\documentclass{elsart}

\usepackage[dvipdf]{graphicx}

\usepackage{amssymb}

\begin{document}

\begin{frontmatter}



\title{170 Nanometer Nuclear Magnetic Resonance Imaging using Magnetic
Resonance Force Microscopy}


\author[ARL]{Kent R. Thurber}
\author[WP]{Lee E. Harrell}
\author[ARL]{Doran D. Smith\corauthref{cor}}
\corauth[cor]{Corresponding author.}
\ead{ddsmith@arl.army.mil}

\address[ARL]{U.S. Army Research Laboratory, Adelphi, MD 20783.}
\address[WP]{Department of Physics, U.S. Military Academy, West Point, New York 10996.}

\begin{abstract}
We demonstrate one-dimensional nuclear magnetic resonance imaging
of the semiconductor GaAs with 170 nanometer slice separation and
resolve two regions of reduced nuclear spin polarization density
separated by only 500 nanometers.  This is achieved by force
detection of the magnetic resonance, Magnetic Resonance Force
Microscopy (MRFM), in combination with optical pumping to increase
the nuclear spin polarization.  Optical pumping of the GaAs
creates spin polarization up to 12 times larger than the thermal
nuclear spin polarization at 5 K and 4 T.  The experiment is
sensitive to sample volumes containing $\sim 4 \times 10^{11}$
$^{71}$Ga$/\sqrt{Hz}$.  These results demonstrate the ability of
force-detected magnetic resonance to apply magnetic resonance
imaging to semiconductor devices and other nanostructures.
\end{abstract}

\end{frontmatter}

\section{Introduction}
\label{Intro} Magnetic Resonance Imaging (MRI) has had many
benefits to medicine and biology.  However, the low sensitivity of
the conventional inductive detection of nuclear magnetic moments
has limited MRI to the micrometer scale and
above\cite{Lee,Pennington,picoliter}.  The alternative technique
of force detection of the magnetic resonance, Magnetic Resonance
Force Microscopy (MRFM)\cite{Sidles,Zuger}, increases both the
sensitivity and the resolution of MRI.  Force detected NMR allows
imaging with resolution well below one micrometer in solids, which
opens up the application of MRI to semiconductor devices, thin
films, and other nanostructures.  In this article, we report force
detection of $^{71,69}$Ga and $^{75}$As magnetic resonance in the
semiconductor GaAs, in combination with optical
pumping\cite{Lampel,Paget} to increase the nuclear spin
polarization. We demonstrate one-dimensional nuclear magnetic
resonance imaging of GaAs with 170 nanometer slice separation and
resolve two regions of reduced nuclear spin polarization density
separated by only 500 nanometers.

\section{Materials and Methods}
\label{materials}

\begin{figure}
\includegraphics[width=3in]{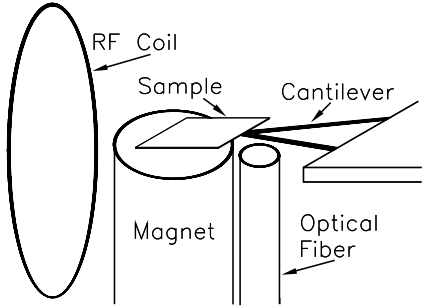}
\caption{Illustration of the MRFM probe head (approximately to
scale). The magnet is iron wire 250 $\mu$m in diameter. The sample
is doped GaAs $260 \times 180 \times 3$ $\mu$m$^3$ and attached to
the Si$_3$N$_4$ cantilever with silver-filled epoxy. The 700
$\mu$m diameter copper RF coil generates the B$_1$ for the
experiment and the single mode optical fiber is used to monitor
the cantilever position.} \label{fig:setup}
\end{figure}

The force measured in a MRFM experiment is the force between two
magnets:  a small ferromagnet, and the nuclear (or electron)
magnetic moments of the sample.  Figure \ref{fig:setup} shows an
illustration of the MRFM probe head.  In our experiment, the
ferromagnet is a 250 $\mu$m diameter iron cylinder. The sample is
a $\sim 260 \times 180 \times 3$ $\mu$m$^{3}$ layer of GaAs, doped
at $0.6 \times 10^{18}$ cm$^{-3}$ Si and $2.0 \times 10^{18}$
cm$^{-3}$ Be.  To detect the force between the sample and the
ferromagnet, the GaAs sample is mounted with silver-filled epoxy
on the end of a microcantilever and positioned 60 $\mu$m from the
surface of the iron magnet. We used a Si$_{3}$N$_{4}$
cantilever\cite{cantilever} coated with 300 \AA\ Ti and 700 \AA\
Au on both sides for thermal conductivity (total spring constant
$k$ of the Au/Ti/Si$_{3}$N$_{4}$/Ti/Au sandwich $\sim 0.05$ N/m).
The loaded cantilever has a mechanical resonant frequency,
$f_{c}=490$ Hz and $Q=75$ at 5 K in He exchange gas. The motion of
the cantilever is observed with a fiber optic
interferometer\cite{interferometer}. Further description of the
cryostat and electronics is given in references
\cite{CaF,Thurber}.

The oscillation of the cantilever is driven by cyclic adiabatic
rapid passage (ARP)\cite{ARP}.  The RF magnetic field
($\omega_{RF}/2\pi = 51.50$ MHz) has a triangle wave frequency
modulation with peak-to-peak frequency width of $2\Omega/2\pi$,
which flips the resonant nuclei at a frequency $f_{ARP}$.  The RF
magnetic field ($2 B_{1} \sim 0.4$ mT) is provided by a 700 $\mu$m
diameter, 1 1/2 turn copper coil. The static magnetic field for
resonance (3.96 T for $^{71}$Ga at 51.50 MHz) is provided by the
combination of an external superconducting magnet and the 250
$\mu$m diameter iron cylinder. The small size of the iron magnet
results in a large magnetic field gradient (6000 T/m) at the
center of the sample which provides the spatial selectivity for
imaging. Only those spins in a total magnetic field satisfying the
resonance condition will contribute to the signal.

Even at 5 K, the thermal spin polarization of the nuclei is rather
small, $6 \times 10^{-4}$. To increase the nuclear spin
polarization, we optically pump the GaAs
sample\cite{Lampel,Paget,Barrett}. An optical fiber shines
circularly polarized light on the sample with a wavelength of 823
nm (near the bandgap of GaAs). Because of the GaAs band structure,
the circularly polarized light creates electron-hole pairs with
the electrons having 50\% net polarization.  The electrons then
polarize the nuclei, through hyperfine interactions primarily at
electronic defects (dynamic nuclear polarization). The optical
pumping was typically done at 0.2 T external applied field because
of the higher nuclear polarization achieved at low field. The
magnetic field was then ramped up for the NMR measurements.
Because of the long $T_{1}$ (21$\pm$5 min)\cite{Thurber}, very
little spin polarization is lost in the roughly one minute
required to change the magnetic field.  The optical power (roughly
1400 W/m$^{2}$) was kept low to avoid heating the sample.
Since the sample is mounted at the end of a thin cantilever,
thermal conductance away from the sample is low ($\approx 25
\mu$W/K).

\section{Results}
\label{results}

\begin{figure}
\includegraphics{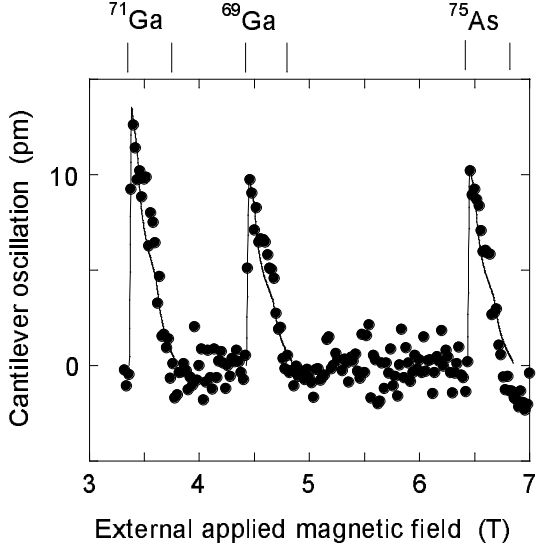}
\caption{Low-resolution 1D images of all three nuclear isotopes of
the GaAs sample, $^{71}$Ga, $^{69}$Ga, and $^{75}$As. Solid line
is the calculated shape of the image. Data taken at $f_{ARP}=33$
Hz (not cantilever mechanical resonance), $2\Omega/2\pi=90$ kHz
with 20 mT step size.} \label{fig:threeisotopes}
\end{figure}

\begin{figure}
\includegraphics{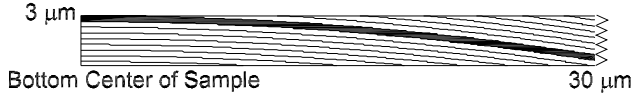}
\caption{Cross-section view of calculated geometry for 330 nm
thick imaging slices.  One slice shaded to correspond to filled
data ($\bullet$) of figure \ref{fig:resolution}.}
\label{fig:slicedraw}
\end{figure}

\begin{figure}
\includegraphics{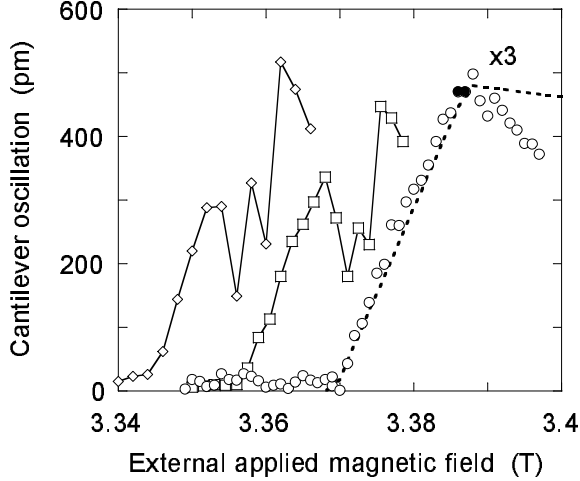}
\caption{High-resolution, optically pumped $^{71}$Ga images. Open
circle data ($\circ$, multiplied by 3) have a 170 nm imaging slice
separation ($2\Omega/2\pi = 4$ kHz, 1 mT step). The dotted line
represents calculations of the imaging slice volume from figure
\ref{fig:slicedraw}, and the two data points corresponding to the
filled slice of figure \ref{fig:slicedraw} are also filled in. The
resolution is demonstrated by destroying the spin polarization in
two closely spaced slices and then imaging, for 670 nm
($\diamond$, offset by 0.03 Tesla) and 500 nm ($\square$, offset
by 0.01 Tesla) separation between the two modified slices. (For
670 nm, $2\Omega/2\pi = 8$ kHz, 2 mT step; For 500 nm,
$2\Omega/2\pi = 8$ kHz, 1.5 mT step.)} \label{fig:resolution}
\end{figure}

\begin{figure}
\includegraphics{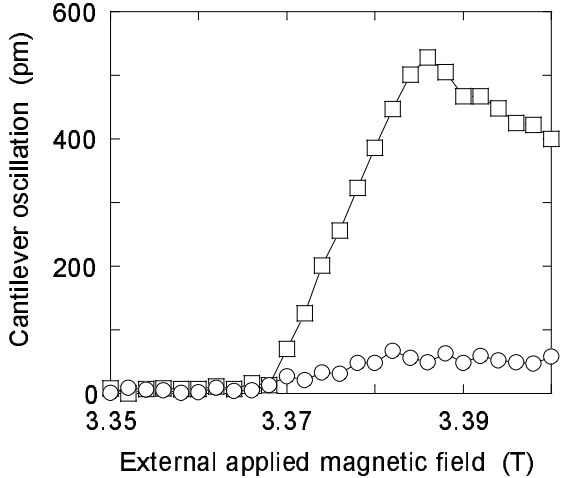}
\caption{Comparison of optically pumped ($\square$) and thermal
($\circ$) $^{71}$Ga nuclear spin polarization ($2\Omega/2\pi = 8$
kHz, 2 mT step).} \label{fig:OP}
\end{figure}

We observed all three naturally abundant nuclear isotopes in GaAs,
$^{71}$Ga, $^{69}$Ga and $^{75}$As, as shown in figure
\ref{fig:threeisotopes}.  The large width of the isotope peaks
reflects the magnetic field gradient and the spatial extent of the
sample.  Each isotope peak is a 1D image of the nuclear spin
polarization density of that isotope.  For our cylindrical magnet,
the imaging slices are shaped like a plate (thin with some
curvature), as shown in cross-section in figure
\ref{fig:slicedraw}.  The first signal at the lowest external
magnetic field is from the bottom center of the sample. The signal
grows rapidly with increasing external field as the slice volume
extends deeper into the 3 $\mu$m thick sample. Once the tip of the
imaging slice extends beyond the sample, the signal size declines
slowly because of the reduced sample volume within the imaging
slice.  Using the observed maximum offset of the signal from the
nominal resonance fields and the saturation magnetization for iron
(2.18 T)\cite{Bozorth}, we can calculate the magnetic field
gradient and the resonant slice geometry. For the imaging slice
corresponding to the peak signal in figure \ref{fig:resolution}
($\bullet$, 3.388 T), the slices are separated by 170 nm with a
total volume of 600 $\mu$m$^3$ and a magnetic field gradient of
6000 T/m.  This volume contains $5 \times 10^{12}$ $^{71}$Ga
nuclei with a signal to noise ratio (SNR) of 14 $/\sqrt{Hz}$.
Figure \ref{fig:resolution} shows single shot measurements of 2.5
or 5 seconds without any averaging. The good agreement between the
leading edge of the data (figure \ref{fig:resolution} $\circ$) and
the calculated signal (dotted line) confirms the calculated
magnetic field gradient. The difference in the decline of the
signal at higher external fields is caused by spatial variation in
the optical pumping, which is most effective at the center of the
sample. Figure \ref{fig:OP} shows the nuclear spin polarization
enhancement by optical pumping. Optical pumping at 0.2 T external
field creates nuclear polarization as much as 12 times greater
than the thermal polarization achieved at 5K and 4 T.

\begin{figure}
\includegraphics{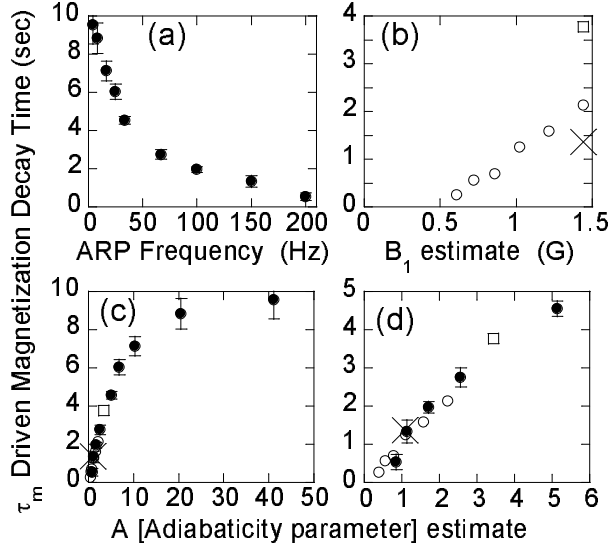}
\caption{(a) $\bullet$ Decay time constant, $\tau_{m}$, of ARP
driven nuclear magnetization as a function of ARP frequency,
$f_{ARP}$. ($^{69}$Ga, $2\Omega/2\pi = 40$ kHz) (b) Decay time as
a function of $B_{1}$ and $\gamma$. ($\circ$ $^{69}$Ga, $\square$
$^{71}$Ga, $\times$ $^{75}$As, $f_{ARP} = 33$ Hz, $2\Omega/2\pi =
94$ kHz) (c) Decay time as a function of adiabatic parameter, $A$.
Data of parts a and b combined. (d) Expanded view of low $A$
region of part c.} \label{fig:ARP}
\end{figure}

To demonstrate our ability to resolve structure in the nuclear
spin polarization density, we want to see how close two planes of
nuclear polarization can be and still be distinguished. To create
contrast in our uniform GaAs sample, we sweep the magnetic field
with constant frequency RF on, which inverts the nuclear
polarization.  During this sweep, we reduce the nuclear
polarization in two closely spaced slices by exposing them to
several seconds of cyclic ARP (see Fig. \ref{fig:ARP}). Between
the two slices of reduced polarization, we leave a third slice
whose polarization is inverted along with the rest of the sample
region, but has nominally unaltered polarization magnitude.
Following this, we measure the resulting polarization. As shown in
figure \ref{fig:resolution}, we can resolve the nuclear spin
polarization signal from two slices separated by only 500 nm. This
is only an upper limit on the resolution of this instrument
because we are doing cyclic ARP on each slice twice: first to
reduce the spin polarization to provide contrast and then a second
time to measure the image. Some of the blurring of the slices
occurs in the creation of the spin-polarization contrast before
imaging.

\section{Discussion}
\label{discussion}

The thickness of the spatial imaging slices in our experiment is
primarily determined by the frequency width of the ARP relative to
the magnetic field gradient.  In this experiment, the highest
resolution data was taken with the peak-to-peak frequency
modulation of the adiabatic rapid passage, $2\Omega/2\pi = 4$ kHz
(equivalent to 0.31 mT for $^{71}$Ga) and the data points were
taken every 1 mT. The data points are thus separated by the
magnetic field step size divided by the field gradient, 1 mT /
(6000 T/m) = 170 nm. We found that reducing the ARP modulation
further rapidly reduced the SNR. This is logical because there are
multiple effects which smear the spatial resolution at the 0.1 mT
level. First, there is $B_{1} \sim 0.2$ mT, which determines the
width of the resonance. Assuming that the adiabatic condition is
met during the passage and that relaxation processes can be
neglected, the modulation of the z-axis magnetization
is\cite{Zuger}
\begin{equation}
M_{z}(t)=M_{0}({\bf r})\frac{\gamma \delta B({\bf r}) - \Omega
t}{\sqrt{[\gamma \delta B({\bf r}) - \Omega t]^{2} + (\gamma
B_{1})^{2}}}
\end{equation}
where $M_{0}({\bf r})$ is the nuclear polarization at position
${\bf r}$, $\gamma$ is the nuclear gyromagnetic ratio ($\gamma =
2\pi \times 13.0$ MHz/T for $^{71}$Ga), $\gamma \delta B({\bf r})
= \gamma B({\bf r}) - \omega_{RF}$ is the offset field from
resonance, and $B({\bf r})$ is the total magnetic field at a
position ${\bf r}$ in the sample. The time variable, $t$, varies
from -1 to +1 during the adiabatic passage. This equation is
correct for the spin 3/2 nuclei in this experiment, $^{71,69}$Ga
and $^{75}$As, as long as the quadrupole coupling is negligible.
In this equation, $B_{1}$ has two effects. Large $B_{1}$ increases
the width of the resonance, thus modulating spins further from the
center of the slice.  Large $B_{1}$ also increases the frequency
width, $\Omega$, required to fully modulate the spins.  A small
$\gamma B_{1}$ relative to $\Omega$ is required to have the slice
width depend primarily on $\Omega$. Having $\gamma B_{1} < \Omega$
provides a large modulation of the spins and sharper edges of the
imaging slice.

However, $B_{1}$ in combination with $\Omega$ and $f_{ARP}$ also
determines the adiabaticity of the rapid passages. To have
adiabatic passages requires the adiabatic parameter $A$,
\begin{equation}
A = \frac{(\gamma B_{1})^{2}}{4\Omega f_{ARP}} \gg 1
\end{equation}
This equation clearly favors large $B_{1}$.  Figure \ref{fig:ARP}
shows the effect of $f_{ARP}$, $B_{1}$, and $\gamma$ on the decay
time $\tau_{m}$ of the nuclear magnetization driven by ARP.  The
available measurement time becomes significant for $B_{1} > 0.05$
mT.  As a result, the experiment requires both $\gamma B_{1}
\lesssim \Omega$ and $(\gamma B_{1})^2 \gg 4\Omega f_{ARP}$.
If we take $\gamma B_{1} = \Omega$, we can simplify these
equations to see that we require roughly $4f_{ARP} \ll \gamma
B_{1} \lesssim \Omega$.  Since minimizing $\Omega$ gives us the
highest resolution, we also want to minimize $f_{ARP}$.  In order
to use the Q enhancement of the mechanical cantilever resonance to
amplify the signal relative to measurement noise, we want $f_{ARP}
= f_{c}$.  In this experiment, a rather large sample was
deliberately used to mass load the cantilever and lower its
resonant frequency, $f_{c}$.

Besides these considerations for the cyclic ARP measurement,
sample properties also can provide limits to the current
experimental resolution.  The intrinsic linewidth of $^{71}$Ga in
GaAs is about 0.2 mT\cite{Barrett}.  As can be seen in figure
\ref{fig:ARP}(c), even if the rapid passage is very adiabatic, the
driven magnetization still decays in about 10 seconds.  Even for
an adiabatic passage, the time spent on resonance is limited by
the spin lock time constant, $T_{1\rho}$.  Another effect which
could limit spatial resolution is spin diffusion. The effect of
spin diffusion was not seen in this experiment.  Spin diffusion
should become important as the resolution is further increased
based on the expected spin diffusion constants, ($D = 10^{-13}$
cm$^{2}$/s for $^{75}$As\cite{Paget}).

The current resolution is limited by the size of $B_{1}$ and
linewidth relative to the magnetic field gradient, not the
sensitivity.  There is room for improvement of the resolution (and
the sensitivity, also) by decreasing the size of the ferromagnetic
particle, which increases the magnetic field gradient.  Higher
field gradients have already been used for ESR
experiments\cite{ESR}.  For detailed comparisons of mechanical
versus inductive detection of magnetic resonance, see references
\cite{Sidles93,Boomerang}.

\section{Conclusions}
\label{conclusions}

We have used force-detected magnetic resonance to image GaAs in
one-dimension with 170 nanometer slice spacing and resolve two
regions of reduced nuclear spin polarization density separated by
only 500 nanometers.  We also demonstrated the combination of
force-detected magnetic resonance with optical pumping to increase
nuclear spin polarization.  We can detect volumes containing $\sim
4 \times 10^{11}$ $^{71}$Ga$/\sqrt{Hz}$ with orders of magnitude
of further improvement expected.  This enables NMR of very small
samples and high resolution imaging.  We envision wide ranging
application of force-detected magnetic resonance to study many
types of samples including biological membranes and molecules,
surfaces and thin films, and semiconductor materials and devices.

\section{Acknowledgements}
\label{ack} This work was partially supported by the DARPA Defense
Science Office Spins in Semiconductors program. The sample was
grown and processed by Peter Newman and Monica Taysing-Lara.  The
authors would like to thank John A. Marohn and John Sidles for
many helpful discussions.



\end{document}